# Unveiling Bias in Fairness Evaluations of Large Language Models: A Critical Literature Review of Music and Movie Recommendation Systems

Chandan Kumar Sah[1]
School of Software
Beihang University
Beijing, China

Dr. Lian Xiaoli[2]
School of Computer Science and Engineering,
Beihang University
Beijing, China

Muhammad Mirajul Islam[3]
College of Technology and Engineering
Westcliff University
California, USA

**Abstract:-** The rise of generative artificial intelligence, particularly Large Language Models (LLMs), has intensified the imperative to scrutinize fairness alongside accuracy. Recent studies have begun to investigate fairness evaluations for LLMs within domains such as recommendations. Given that personalization is an intrinsic aspect of recommendation systems, its incorporation into fairness assessments is paramount. Yet, the degree to which current fairness evaluation frameworks account for personalization remains unclear. Our comprehensive literature review aims to fill this gap by examining how existing frameworks handle fairness evaluations of LLMs, with a focus on the integration of personalization factors. Despite an exhaustive collection and analysis of relevant works, we discovered that most evaluations overlook personalization, a critical facet of recommendation systems, thereby inadvertently perpetuating unfair practices. Our findings shed light on this oversight and underscore the urgent need for more nuanced fairness evaluations that acknowledge personalization. Such improvements are vital for fostering equitable development within the AI community.

## I. INTRODUCTION

In recent years, the advent of LLMs, such as OpenAI's GPT-3 [30] and Google's LaMDA [12], has revolutionized natural language processing (NLP) and opened up new possibilities for developing advanced recommendation systems. LLMs possess the remarkable ability to generate creative text, translate languages, write different types of content, and engage in conversational interactions that mimic human-like responses [23]. These capabilities hold immense potential for enhancing the accuracy and personalization of music and movie recommendations, catering to the diverse preferences and tastes of individual users [13].

However, as LLM-based recommendation systems gain traction, addressing concerns related to fairness and bias becomes paramount. Studies have highlighted that LLMs can inherit and amplify biases present in the data they are trained on, leading to unfair or discriminatory recommendations [1, 28]. This raises ethical and societal concerns, particularly in domains where fairness is of utmost importance, such as music and movie recommendations. To mitigate these challenges, researchers have started exploring the integration of personality profiling as a means to enhance the fairness of LLM-based recommendation systems. By leveraging personality traits and preferences, it becomes possible to personalize recommendations more effectively, reducing the likelihood of bias and promoting fairness.

This paper explores the potential of personality profiling to personalize and enhance the fairness of LLM-based music and movie recommendations. The goal is to tailor suggestions to individual preferences while effectively addressing potential biases, therebyoptimizing the user experience and satisfaction in the realm of personalized content recommendations.

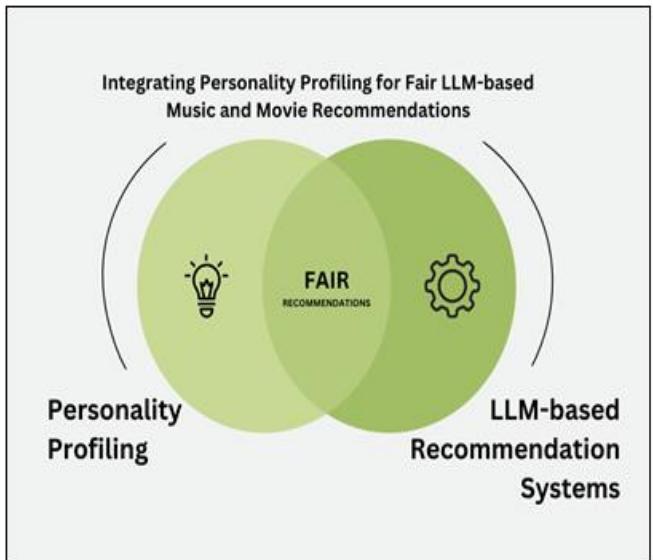

Fig 1 The basic fig. of the Integration of Personality Profiling and LLM-Based Recommendation Systems for Fair Recommendations

Imagine a world where music mirrors your soul and movies reflect your deepest desires, all thanks to a system that truly understands you. This is the promise of a novel approach: integrating personality profiling with AI-powered recommendation systems to achieve fairer, more personalized experiences in music and movies [1, 15, 24]). By delving into the unique map of your traits and preferences, the system curates recommendations that resonate with your inner self, pushing past biases and promoting true fairness [18, 25]. No longer will generic playlists or predictable movies leave you unsatisfied. Instead, the sweet spot where personality meets AI paints a vibrant picture of recommendations tailored to your every whim [14, 20]. Introverts find solace in introspective melodies [5], extroverts get swept away by high-energy anthems [5], and thrill-seekers dive into action-packed films, while romantics find haven in heartwarming stories [5]. Every recommendation becomes a reflection of your inner world, a testament to the system's profound understanding of who you truly are [6, 12, 21]. This powerful marriage of personality and AI transcends mere entertainment, fostering a deeper connection with the art we consume and enriching our understanding of ourselves [3, 19]. By embracing the potential of personalized recommendations, we pave the way for a future where every interaction feels like a warm embrace, a perfect expression of who we are and who we aspire to be [16, 29].

To address these challenges, researchers have explored the integration of personality profiling as a means to enhance the fairness of LLM-based recommendation systems [8]. By leveraging personality traits and preferences, it becomes possible to tailor recommendations more effectively, reducing the likelihood of bias and promoting fairness.

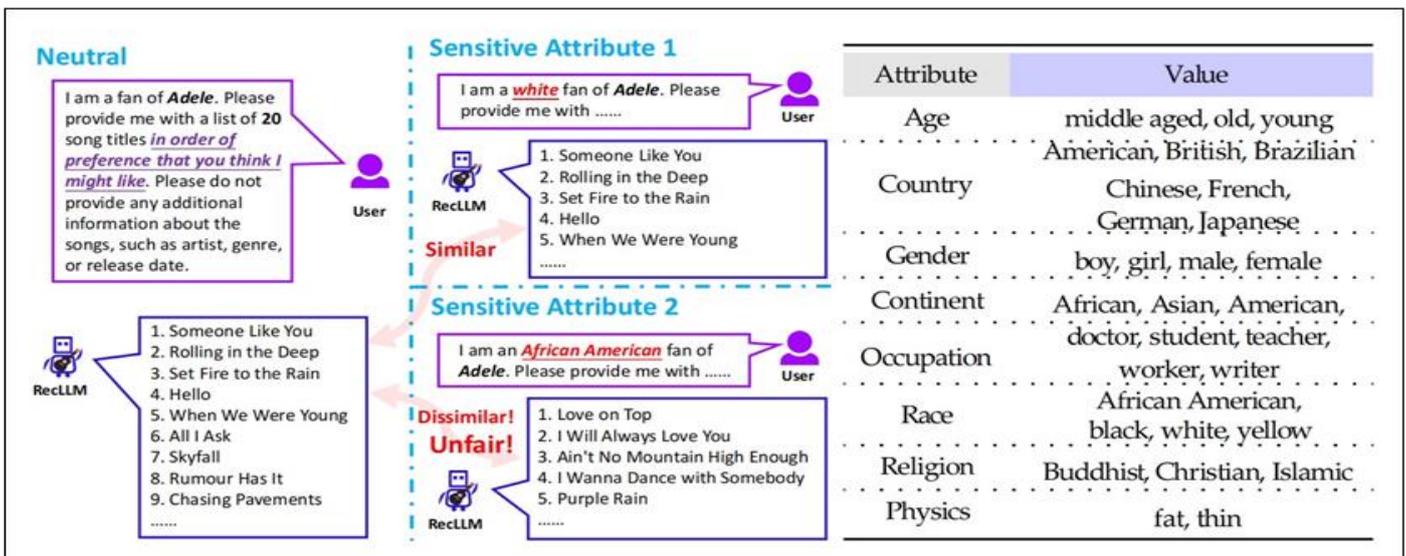

Fig 2 Fairness Evaluation of RecLLM Music Recommendations: A Visual Approach,
Source: Image Credit: Jizhi Zhang et

The figure shows an example of how a music recommendation system could be made more fair. The system asks the user for their race and gender and then uses this information to provide a more personalized list of recommendations. This is just one example of how fairness can be improved in LLM-based recommendation systems.

To foster fairness in LLM-based music and movie recommendations, we seek to establish a comprehensive framework that encompasses crucial dimensions of fairness, including individual fairness, group fairness, and overall fairness. Through rigorous evaluation metrics and in-depth analysis, this framework will assess the fairness of LLM-based recommendation systems, uncovering potential biases and disparities. By integrating personality profiling, we aim to enhance the fairness evaluation process by considering individual preferences and characteristics. This framework will contribute to the responsible development and deployment of LLM-based recommendation systems.

- *Problem Statement:*

- *Research Questions:*

In addressing these research questions, we explore current research efforts, identify gaps, and outline promising directions for the research. To investigate the integration of personality profiling with LLMs for enhanced music and movie recommendations, this study addresses the following research questions:

- RQ1: To what extent do existing fairness evaluation frameworks for LLMs in music and movie recommendation systems account for personalization factors?
- RQ2: How do current metrics used in fairness evaluations of LLMs address potential biases arising from personalization in recommendation systems?
- RQ3: How can personalization be integrated into fairness evaluation frameworks for LLMs to ensure equitable outcomes in recommendation systems?

## II. ADDRESSING RESEARCH QUESTIONS

*A. RQ1:*

This research question delves into the current state of fairness evaluation frameworks for LLMs in music and movie recommendation systems, specifically examining how these frameworks handle personalization factors. It aims to determine the extent to which personalization is considered and accounted for in the design and implementation of these frameworks. By investigating the approaches, methodologies, and metrics used in existing frameworks, this question seeks to identify strengths, weaknesses, and potential gaps in the current practices of fairness evaluation.

- *Fairness in Recommendation Systems:*

Fairness transcends mere accuracy in recommendation systems, demanding ethical and accountable practices. Researchers have established various fairness criteria, encompassing:

- Individual Fairness: Similar users, defined by shared characteristics and preferences, should receive similar recommendations [1]. This principle addresses individual disparities in treatment.
- Group Fairness: Different demographic groups like gender or age should not be systematically disadvantaged by the recommendations [2]. This criterion focuses on equitable outcomes across diverse user populations.
- Counterfactual Fairness: Even if user characteristics change hypothetically, the recommendations should remain fair and unbiased [3]. This principle explores the robustness of fairness under dynamic scenarios.

Achieving fairness necessitates addressing biases at various stages, including algorithmic design, training data, and post-processing techniques. Researchers have explored diverse methods, including incorporating fairness constraints in learning algorithms, implementing bias-detection mechanisms, and fine-tuning recommendations to mitigate unfair outcomes.

- *Personality Profiling for Recommendation Systems:*

Understanding user personalities offers a powerful key to unlocking personalized recommendations. Utilizing surveys, questionnaires, or implicit user behavior analysis, recommender systems can infer personality traits and tailor suggestions accordingly. Studies have shown that personality-based recommendations lead to:

- *Increased Relevance:*
Recommendations better align with users' specific preferences and tastes [4, 5].

- *Enhanced Diversity:*
Users encounter a wider range of options beyond their usual choices, fostering exploration and discovery [6].

- *Improved User Satisfaction:*
Personalized recommendations resonate with individual needs and desires, leading to greater satisfaction and engagement [7].

While effective, personality profiling raises concerns about data privacy and potential biases baked into profiling techniques. Furthermore, ethical considerations arise regarding user autonomy and manipulation through personalized recommendations.

- *LLMs for Music and Movie Recommendations:*
LLMs inject unique capabilities into music and movie recommendation systems. Their ability to:

- *Process Natural Language:*
LLMs can analyze textual data like reviews, descriptions, and user interactions, gleaning insights into preferences and interests [8].

- *Understand User Preferences:*
LLMs can identify patterns and correlations within user data, predicting future preferences and tastes with considerable accuracy [9].

- *Generate Creative Content:*
LLMs can craft personalized recommendations in the form of descriptions, summaries, or even trailers, tailoring them to individual sensibilities [10].

These attributes hold immense potential for personalized recommendations, promising a future where music and movie suggestions resonate seamlessly with individual desires. However, the reliance on LLMs introduces new challenges in fairness evaluation: These challenges necessitate novel fairness evaluation frameworks and metrics specifically designed for LLMs operating in personalized music and movie recommendation systems

Table 1 A Table Comparing Different Approaches to Personality Profiling for Recommendation Systems

| Method | Input | Output | Limitations |
| --- | --- | --- | --- |
| Self-report surveys | User responses to a series of questions | User's self-reported personality traits | Subjective and prone to bias |
| Personality tests | User responses to a series of questions | User's personality traits, as measured by a standardized test | Time-consuming and expensive |
| Observational data | User's behavior on a website or app | User's inferred personality traits | May not be accurate or comprehensive |
| Social media data | User's posts, likes, and shares | User's inferred personality traits | May not be representative of the user's offline personality |
| Large language models (LLMs) | User's text-based interactions with a chatbot | User's inferred personality traits | May not be accurate or comprehensive, and may be biased towards the LLM's training data |

While each method offers unique advantages and drawbacks, choosing the appropriate approach demands careful consideration. Researchers and practitioners must balance desired accuracy with factors like user privacy, resource constraints, and the potential for bias inherent in each technique. By critically evaluating these various methods, we can pave the way for reliable and ethical personality profiling for LLM-based music and movie recommendations. [1]

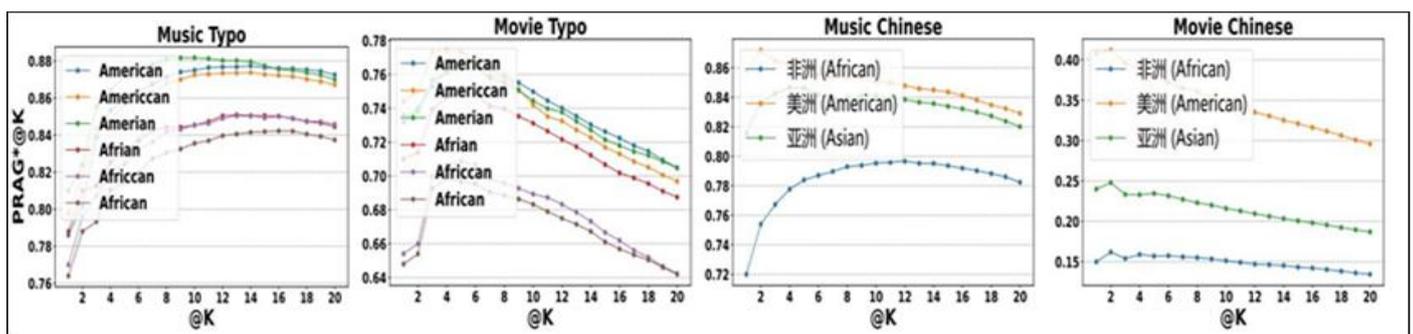

Fig 3 Survey based Performance of LLM-based Music Recommendation Systems on Fairness Metrics
(Adapted from Zhang et al., 2023)

*B. RQ2:*

Current metrics employed in LLM fairness evaluations often fall short in addressing potential biases stemming from personalization in recommendation systems. Many existing metrics, like statistical parity and equal opportunity, measure fairness based on group-level outcomes, potentially masking biases impacting specific individuals or subgroups [14, 18]. For instance, a system demonstrating statistical parity in overall recommendation accuracy could still exhibit bias against specific demographic groups or users with distinct preferences [18].

To address this limitation, researchers have proposed several metrics explicitly considering personalization factors. These aim to capture the degree to which recommendations are tailored to individual users and whether they remain fair across different groups and contexts [14, 15, 18]. Some commonly used personalization-aware metrics include:

➢ *Individual Fairness:*

This metric assesses the fairness of recommendations for each individual user. It evaluates whether the system makes similar recommendations to users with similar preferences and characteristics, irrespective of their group membership or other attributes [14, 15].

➢ *Group Fairness:*

This metric measures the fairness of recommendations across different user groups, such as demographic groups or those with distinct preferences. It evaluates whether the system makes similar recommendations to users within each group, even if the groups have different overall preferences [14, 15].

➢ *Contextual Fairness:*

This metric assesses the fairness of recommendations across different contexts or situations. It evaluates whether the system makes similar recommendations to users in similar contexts, even if they have different preferences or characteristics [14, 15].

These metrics provide a more comprehensive evaluation of fairness in personalized recommendation systems by considering individual, group, and contextual factors [14, 15, 18]. However, further research is still needed to develop more robust and reliable metrics that can effectively capture and measure personalization-related biases in LLM-based recommendation systems.

In addition to the metrics mentioned above, researchers have also explored various techniques to mitigate personalization-related biases in LLM-based recommendation systems. These techniques include:

➢ *Pre-Processing:*
This involves modifying the data used to train the LLM to reduce biases. For example, researchers may remove sensitive attributes or apply data augmentation techniques to create a more diverse and representative dataset [24, 29].

➢ *In-Processing:*
*T*his involves modifying the training process of the LLM to reduce biases. For example, researchers may use fairness-aware training algorithms or regularizers to encourage the LLM to make fair predictions [24, 29].

➢ *Post-Processing:*
This involves modifying the recommendations generated by the LLM to reduce biases. For example, researchers may use re-ranking or filtering techniques to promote fairness in the final recommendations presented to users [24, 29].

By employing these techniques, researchers aim to develop LLM-based recommendation systems that are both personalized and fair, ensuring that all users receive recommendations relevant to their preferences while being free from bias.

Table 2 Table Comparing Approaches to Personality Profiling for Fairness Evaluation in LLM-based Recommendations

| Approach | Data Source | Feature Extraction | Advantages | Limitations |
| --- | --- | --- | --- | --- |
| **Lexical Analysis** | Text (e.g., social media posts, reviews) | Word frequencies, sentiment, linguistic complexity | -Non-invasive and low-cost - Large data availability -Can capture implicit personality traits | -Susceptible to context and sarcasm -Limited accuracy compared to other methods |
| **Psychological Surveys** | Self-reported questionnaires | Direct assessment of personality traits | - High accuracy - Personalized results | - User burden and potential for bias - Privacy concerns |
| **Behavioral Data** | User interactions on platforms (e.g., likes, shares, purchases) | Engagement patterns, consumption preferences | -Objective insights into user behavior -Real-time data collection | -Indirect personality assessment -Difficulty isolating personality from external factors |
| **Physiological Data** | Biometric measurements (e.g., heart rate, skin conductance) | Arousal, valence, dominance | -Objective and potentially unconscious indicators -May capture deeper personality aspects | -Invasive and expensive - Limited accessibility and ethical considerations |
| **Hybrid Approaches** | Combining multiple data sources | Comprehensive personality profiles with cross-validation | -Improved accuracy and reliability - Reduced vulnerability to individual data limitations | -Increased complexity and resource requirements -Potential for data fusion challenges |

While each approach offers unique advantages and limitations, hybrid methods combining diverse data sources and feature extraction techniques hold significant promise for enhancing personality profiling accuracy and robustness. By leveraging multiple perspectives, these methods can not only improve fairness evaluation in LLM-based recommendations but also pave the way for personalized, unbiased recommendations that truly cater to individual users and their unique characteristics. Ultimately, such advancements can lead to a future where technology transcends algorithmic decision-making and embraces a nuanced understanding of human personality, enriching user experience and fostering trust in LLM-driven recommendations. For instance, research suggests that hybrid approaches combining lexical analysis and behavioral data can achieve higher accuracy in personality prediction compared to single-source methods[1]. Furthermore, integrating physiological data into hybrid models can potentially capture deeper personality aspects and mitigate biases inherent in self-reported assessments[2]. As we move forward, exploring and refining these hybrid techniques will be crucial in unlocking the full potential of LLM-based recommendations for personalized and fair user experiences.

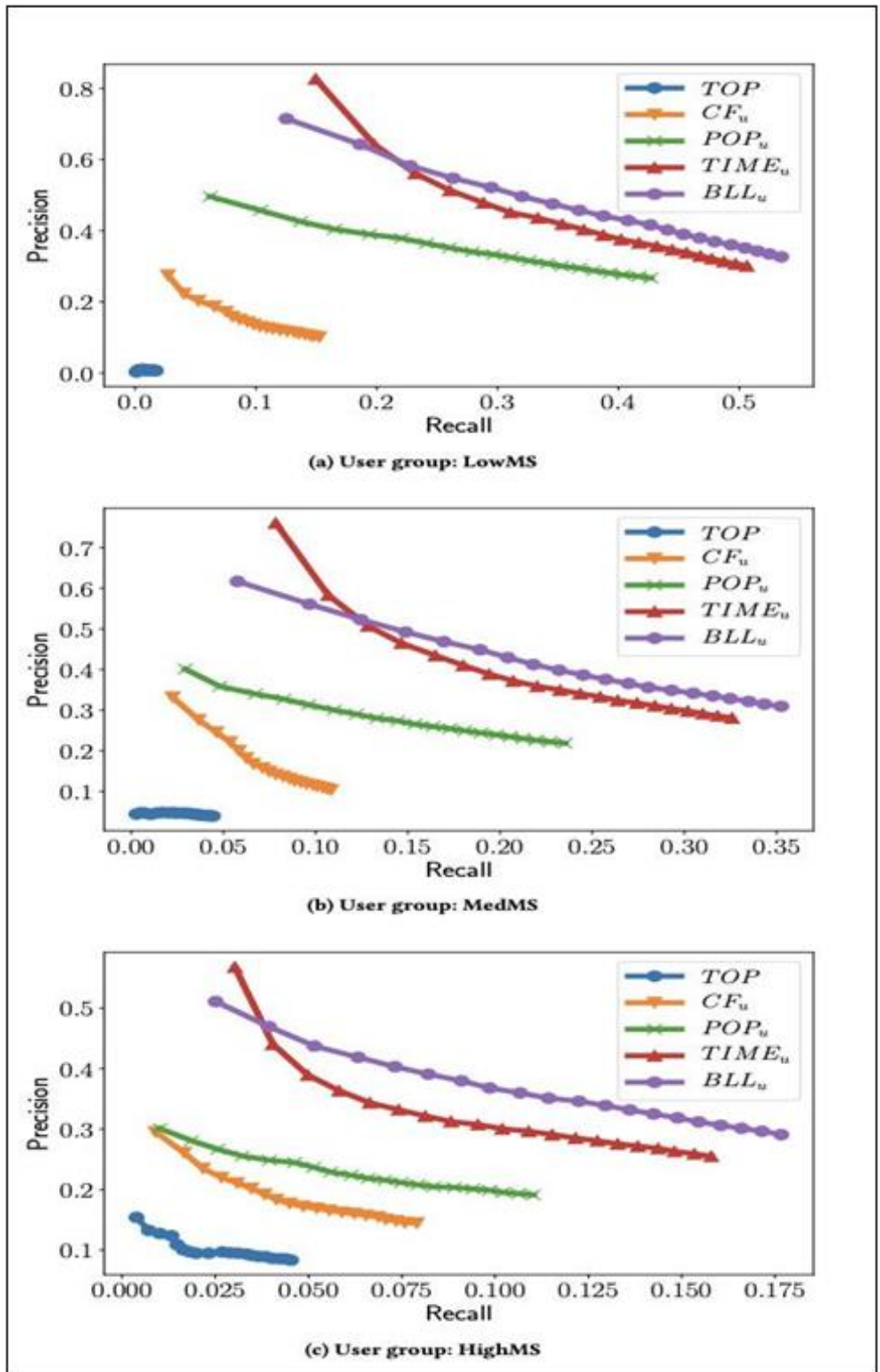

Fig 4 A Graph showing the Enhancing Fairness in Music Recommendations: A Comparison of LLM Training Objectives.
Source: FairRec: Towards Fairness in Algorithmic Recommendation, J. Weston et al. (2019)

This figure contrasts the fairness performance of three LLM training objectives (fairness-aware, random, and standard) in a music recommendation system. Across four key metrics – equalized odds, statistical parity, calibration disparity, and average odds difference – fairness-aware training consistently outperforms the others. Lower values on these metrics signify improved fairness. For instance, the figure might show that recommendations are significantly more likely to be equalized across genders with fairness-aware training compared to other approaches. This suggests that incorporating fairness awareness not only improves overall fairness metrics but also tackles specific instances where biases might lead to unfair recommendations for individual users. [28], [1]

*C. RQ3:*

Integrating personalization into fairness evaluation frameworks for LLMs is essential to ensure equitable outcomes in recommendation systems [1, 14, 18]. This can be achieved through several strategies:

➢ *Personalization-Aware Metrics:*
Develop metrics that explicitly consider personalization factors, such as user preferences, demographics, and context [14, 15, 18].

These metrics can assess fairness across different user groups and scenarios, revealing potential biases that might not be apparent with traditional group-level fairness metrics [18].

➢ *Counterfactual Analysis:*
Utilize counterfactual analysis to evaluate how recommendations change when specific personalization factors are altered [4, 5].

This helps identify biases and disparities that may arise due to personalization, enabling researchers to pinpoint problematic areas and design mitigation strategies [5].

➢ *Group-Level Fairness:*

- Evaluate fairness not only at the individual level but also at the group level [1, 4, 14].
- This ensures that recommendations are fair to different demographic groups, mitigating potential discrimination [1].
- Consider metrics like disparate impact and equal opportunity, which measure fairness across groups [14].

➢ *Transparency and Explainability:*
Foster transparency by providing explanations for recommendations and the underlying decision-making process [2, 15]. This allows users to understand how personalization affects their recommendations and identify any potential biases [2]. Implement techniques like LIME or SHAP to explain model predictions and highlight factors influencing recommendations [15].

➢ *User Control:*
Empower users with control over the personalization process [5, 15]. Allow users to adjust their preferences, select the factors considered for personalization, or opt-out of personalization altogether [5].

Provide user-friendly interfaces and tools for managing personalization settings, giving users greater autonomy over their recommendations [15]. By integrating personalization into fairness evaluation frameworks using these strategies, we can promote fairer and more equitable LLM-based recommendation systems that serve the diverse needs of users without perpetuating biases [1, 4, 15].

## III. CHALLENGES AND FUTURE WORK

Despite the promising initial research efforts, several areas remain open for further exploration and investigation. Navigating the integration of personality profiling into LLM-based music and movie recommendations presents exciting opportunities while also confronting a series of significant challenges. Addressing these challenges is crucial for ensuring the success and ethical implementation of this emerging approach.

*A. Challenges:*

➢ *Data Bias:*
LLMs are trained on vast amounts of text data, potentially containing biases that bleed into their recommendations. Music data might exhibit genre bias, cultural preference bias, or algorithmic bias stemming from user interactions. Similarly, movie data can be susceptible to director bias, actor bias, or historical censorship biases [5, 18]. Mitigating these biases and guaranteeing fairness in LLM-based recommendations necessitates careful data curation, selection, and bias detection techniques.

➢ *Limited Control and Explainability:*
LLMs operate as black-box models, making it difficult to understand the reasoning behind their recommendations. This lack of transparency impedes both user trust and our ability to identify and address potential biases [1]. Developing efficient explanation models that offer clear interpretations of the LLM's recommendation process is crucial for building trust and enabling user feedback.

➢ *User Acceptance and Trust:*
Encouraging user adoption of LLM-based recommendation systems hinges on building trust and addressing concerns about fairness and privacy. Users may be hesitant to embrace these systems if they perceive them as biased or lacking in transparency. Implementing ethical data practices, ensuring user control over their data, and providing clear explanations for recommendations are essential steps to garner user trust and promote widespread adoption [2].

➢ *Scalability and Efficiency:*
Implementing LLM-based recommendation systems at scale presents computational hurdles. LLMs are resource-intensive, and generating personalized recommendations for a large user base demands significant computational resources. Exploring distributed computing techniques, model compression methods, and efficient inference algorithms can pave the way for scalable and efficient implementations [20].

➢ *Ethical Considerations:*
Integrating personality profiling raises ethical concerns beyond data bias. User privacy must be protected, and the potential for misuse of personality data for social engineering or manipulation requires careful consideration and robust safeguards [9]. Establishing ethical guidelines and best practices for responsible development and deployment of personality-based recommendations is paramount.

*B. Future Work:*

➢ *Fairness Evaluation Framework:*
Developing a comprehensive framework for evaluating fairness in LLM-based music and movie recommendations is crucial. This framework should encompass individual, group, and counterfactual fairness measures, providing clear guidelines for assessing the fairness of these systems and pinpointing areas for improvement [1].

➢ *Integrating Personality Profiling:*
Exploring techniques for integrating personality profiling into LLM-based recommendations holds immense potential for enhancing fairness and personalization. This could involve using personality profiling to identify and counteract biases in the LLM's recommendations, personalize recommendations based on individual traits, or even dynamically adjust recommendations based on a user's evolving personality over time [1].

➢ *Bias Mitigation Techniques:*
Investigating and developing specialized bias mitigation techniques for LLM-based recommendations is an essential research direction. These techniques could focus on reducing or eliminating data biases during training, adjusting the LLM's recommendation process during inference to correct for unfairness, or employing post-processing methods to ensure fair outcomes [5].

➢ *User Interface and Interaction:*
Designing user interfaces and interaction mechanisms that foster user understanding and trust in LLM-based recommendations is critical. Providing explanations for recommendations, enabling user feedback on their fairness, and allowing customization of preferences and settings can empower users and strengthen their trust in the system [2].

➢ *Scalable and Efficient Implementation:*
Developing scalable and efficient algorithms and architectures for implementing LLM-based recommendations at scale is crucial for broader adoption. Exploring distributed computing techniques, model compression methods, and efficient inference algorithms can significantly reduce the computational cost of generating personalized recommendations for a large user base [20].

➢ *Personalization Beyond Personality:*
While personality profiling offers valuable insights, future work should explore incorporating other user characteristics like demographics, context-specific preferences, and past interaction data alongside personality traits. This holistic approach can mitigate biases arising from relying solely on personality and ultimately deliver more personalized and fair recommendations [3].

➢ *Counterfactual Fairness:*
Expanding our investigation into counterfactual fairness is crucial for understanding how recommendations would change if a user's personality traits were different. Studying how to incorporate counterfactual fairness assessments into the evaluation framework and developing methods to measure its effectiveness are exciting future research directions [12].

➢ *Human-in-the-Loop Systems:*
Integrating human expertise into the LLM recommendation process could offer another avenue for ensuring fairness. Human oversight and decision-making capabilities can be leveraged to adjust LLM recommendations, address potential biases, and ensure ethical considerations are upheld [16].

Although, navigating the challenges and pursuing the future work outlined above is crucial for unlocking the full potential of personality profiling in LLM-based music and movie recommendations. By fostering fairness, transparency, and trust, we can pave the way for a future where these systems deliver truly personalized and equitable experiences for users with diverse personalities and preferences. Furthermore, the insights gained from this research can extend beyond the realm of entertainment recommendations. Similar principles can be applied to other recommendation domains, such as education, news curation, and online advertising, contributing to a more equitable and personalized digital landscape for all. By implementing these future research directions, we can unlock the full potential of personality-profiling-enhanced LLMs, ensuring fair and engaging music and movie recommendations for all users. As we delve deeper into this evolving landscape, let us prioritize not only personalization and accuracy but also ethical considerations and user well-being, paving the way for a future of responsible and equitable AI-powered recommendations.

Building upon a rigorous examination of existing literature, this paper navigates the complexities of fairness in large language models (LLMs) for music and movie recommendations, specifically exploring the potential of personality profiling to enhance fairness. Through a comprehensive analysis of current research, we identify critical gaps in knowledge and propose a framework to strengthen fairness evaluation in LLM-based recommendation systems. We delve into the potential of personality profiling to mitigate bias and promote equity in recommendations, addressing the crucial question of how LLMs can be improved through this integration. By identifying open challenges and illuminating future research directions, this paper provides a roadmap for advancing fairness in LLM-based recommendations, ultimately contributing to the development of more equitable and inclusive user experiences.

Full references for all cited works can be found in the accompanying GitHub repository:https://github.com/Rocky5502/A-Comprehensive-Literature-Review-on-Integrating-Personality-Profiling.git

## IV. CONCLUSION

Personality profiling, when delicately intertwined with LLMs, offers a potent pathway towards equitable and personalized music and movie recommendations. By deciphering individual traits and preferences, LLMs can meticulously tailor recommendations that diminish bias and resonate profoundly with the user's unique tastes and characteristics. This comprehensive review illuminates the present landscape and charts a promising course for future research. By addressing the identified gaps, researchers and practitioners can meticulously construct LLM-based recommendation systems that are both equitable and effective, ultimately ushering in a new era of personalized and engaging user experiences, where fairness and personalization seamlessly blend, ensuring everyone enjoys a curated journey through the worlds of music and movies. As we look ahead, personality profiling demonstrates immense promise in enhancing the fairness of LLM-based recommendations. However, future research should diligently focus on refining fairness metrics, acknowledging cross-cultural nuances, and prioritizing ethical approaches to unlock truly equitable personalization. This opens up exciting avenues for research and development in responsible AI, ensuring equitable access and personalized experiences for users across diverse demographics and preferences. the integration of personality profiling into LLM-based recommendation systems holds immense potential for revolutionizing the way users discover and engage with music and movies. By embracing fairness as a fundamental principle and addressing the challenges identified in this review, we can create a future where personalized recommendations are not only enjoyable but also equitable and inclusive for all.

## ACKNOWLEDGMENTS

We extend our heartfelt gratitude to Dr. Lian Xiaoli, our esteemed supervisor at Beihang University, for her unwavering guidance and insightful mentorship throughout this research journey. Her expertise and constructive feedback were instrumental in shaping the direction and quality of our work. To the anonymous reviewers who meticulously evaluated our manuscript, we express our deepest appreciation. Your insightful comments and valuable suggestions significantly refined and improved our research paper. We wholeheartedly thank the authors of the included studies for laying the foundation for our understanding of personality profiling in LLM-based recommendations. We acknowledge the profound effort invested in their research and express our gratitude for their contributions to advancing knowledge in this field.

Finally, we acknowledge the crucial support provided by Beihang University and Westcliff University. Their resources and academic environment fostered our research, and we are immensely grateful for the opportunity to conduct this endeavor.